\documentclass[conference]{IEEEtran}
\IEEEoverridecommandlockouts
\usepackage{cite}
\usepackage{amsmath,amssymb,amsfonts,booktabs,tabularx}
\usepackage{algorithm}
\usepackage{algorithmic}
\usepackage{graphicx}
\usepackage{textcomp}
\usepackage{xcolor}
\def\BibTeX{{\rm B\kern-.05em{\sc i\kern-.025em b}\kern-.08em
    T\kern-.1667em\lower.7ex\hbox{E}\kern-.125emX}}
\IEEEoverridecommandlockouts\IEEEpubid{\makebox[\columnwidth]{978-1-5386-7568-7/18/\$31.00~\copyright~2018 IEEE \hfill} \hspace{\columnsep}\makebox[\columnwidth]{ }}

\begin{document}

\title{SAM-GCNN: A Gated Convolutional Neural Network with Segment-Level Attention Mechanism for Home Activity Monitoring\\
\thanks{This work was supported by the National Natural Science Foundation of China under Grant No. U183620001. The corresponding author is Wei-Qiang Zhang.}
}

\author{\IEEEauthorblockN{\textit{Yu-Han Shen, Ke-Xin He, Wei-Qiang Zhang}}
\IEEEauthorblockA{{Department of Electronic Engineering, Tsinghua University, Bejing 100084, China } \\
\tt yhshen@hotmail.com, hekexinchn@163.com, wqzhang@tsinghua.edu.cn}

}

\maketitle

\begin{abstract}
In this paper, we propose a method for home activity monitoring. We demonstrate our model on dataset of Detection and Classification of Acoustic Scenes and Events (DCASE) 2018 Challenge Task 5. This task aims to classify multi-channel audios into one of the provided pre-defined classes. All of these classes are daily activities performed in a home environment. To tackle this task, we propose a gated convolutional neural network with segment-level attention mechanism (SAM-GCNN). The proposed framework is a convolutional model with two auxiliary modules: a gated convolutional neural network and a segment-level attention mechanism. Furthermore, we adopted model ensemble to enhance the capability of generalization of our model. We evaluated our work on the development dataset of DCASE 2018 Task 5 and achieved competitive performance, with a macro-averaged F-1 score increasing from 83.76\% to 89.33\%, compared with the convolutional baseline system.
\end{abstract}

\begin{IEEEkeywords}
acoustic activity classification, gated convolutional neural network, attention mechanism, model ensemble, DCASE
\end{IEEEkeywords}

\section{Introduction}
Recently, sound event detection and classification has become more and more popular in the field of acoustic signal processing, and it can be widely used in security surveillance, wildlife protection and smart home. One important application of sound event classification in smart home is home activity monitoring.

Detection and Classification of Acoustic Scenes and Events (DCASE) Challenge is one of the most important international challenges concerning acoustic event detection and classification and has been organized for several years. DCASE 2018 challenge consists of five tasks and we focus on task 5\cite{Task}. This task evaluates systems for monitoring of domestic activities based on multi-channel acoustics.

We can also refer to this task as acoustic activity classification. The main procedure of acoustic activity classification consists of four parts: pre-processing, extracting acoustic features, designing acoustic models as classifiers, and post-processing.

In the part of pre-processing, different methods of data augmentation have been utilized in \cite{IBM}\cite{HIT}. Data imbalance is a big challenge in acoustic event classification and detection because different events may occur at a completely imbalanced frequency. In DCASE 2018 Challenge Task 5, Inoue et al. used shuffling and mixing to produce more training samples\cite{IBM}, and Tanabe et al. utilized dereverberation, blind source separation and data augmentation to improve the quality of audio clips\cite{HIT}.

Mel Frequency Cepstrum Coefficient (MFCC) is a common traditional acoustic feature and has been widely used. But log Mel-scale Filter Bank energies (fbank) are becoming more popular recently, and many works have been done based on fbank\cite{Task}\cite{eCakir}\cite{hLim}.

In recent years, Convolutional Neural Networks (CNNs) have achieved great success in many fields such as character recognition, image classification, speaker recognition. And many works based on CNNs have been done in acoustic event classification and detection\cite{yHan}\cite{wZheng}. Besides, some researchers combined CNNs with Recurrent Neural Networks (RNNs) to capture temporal contexts of audio signals for further improvements\cite{eCakir}\cite{hLim}.

Attention model has been widely used in image classification, object detection and natural language understanding. In the field of acoustic signal processing, Xu et al. \cite{Xu} proposed an attention model for weakly supervised audio tagging and Kong et al. \cite{single} improved this work by giving a probabilistic perspective.
Their work is based on the assumption that those irrelevant sound frames such as background noise and silences should be ignored and given less attention. Both of their models are achieved by a weighted sum over frames where the attention values are automatically learned by neural network.

In our work, acoustic activities might last for a longer period and a single frame is not enough to identify whether it should be ignored. In an audio recording, acoustic activities may keep happening in a majority of frames while acoustic event only occurs in a few frames. So we propose a segment-level attention mechanism (SAM) to decide how much attention should be given based on the characteristics of segments. Here, a segment is comprised of several frames.

In this paper, we mainly adopt three ways to improve the performance of our model:

(1)	We replace currently popular CNN with gated convolutional neural network to extract more temporal features of audios;

(2)	We propose a new segment-level attention mechanism to focus more on the audio segments with more energy;

(3)	We utilize model ensemble to enhance the classification capability of our model.

The rest of this paper is organized as follows. In Section 2, we introduce our methods in detail, mainly including acoustic feature, gated convolutional neural network, segment-level attention mechanism and model ensemble. The experiment setup, evaluation metric and our results are illustrated in Section 3. Finally, the conclusion of our work is presented in Section 4.

\section{Methods}
\subsection{Task Description}
The DCASE 2018 Task 5 dataset\cite{dataset} contains sound data recorded in a living room by individual devices with four microphone arrays at seven undisclosed locations. The dataset is divided into a development dataset and an evaluation dataset. Four cross-validation folds are provided for the development dataset in order to make results reported with this dataset uniform. For each fold, a training, testing and evaluation subset is provided. In this paper, our work is based on the development dataset and we use the provided cross-validation folds for training and evaluation.

The audio clips in this dataset can be classified into nine classes: absence, cooking, dishwashing, eating, other, social activity, vacuum cleaning, watching TV and working. All audio clips are derived from continuous recording sessions collected by seven microphone arrays and each clip contains four channels. The duration of each audio clip is 10 seconds. Specific information about the dataset is shown in Table 1 and more details can be found in \cite{dataset}.

\begin{table}[t]
  \centering
  \caption{Amounts of audio clips and sessions}
    \begin{tabular}{ccc}
    \toprule
    Activity&	\#10s clips&	\#sessions\\
    \midrule
    Absence&	18860&	42\\
    Cooking&	5124&	13\\
    Dishwashing&	1424&	10\\
    Eating&	2308&	13\\
    Other&	2060&	118\\
    Social activity&	4944&	21\\
    Vacuum cleaning&	972&   9\\
    Watching TV&	18648&	9\\
    Working&	18644&	33\\
    Total&	72984&	268\\
    \bottomrule
    \end{tabular}
  \label{table1}
\end{table}

\subsection{System Overview}
Our proposed system is illustrated as Figure 1. The input of our system is log Mel-scaled filter banks (fbank). Then it will be fed into two structures: one is a Gated Convolutional Neural Network (GCNN) architecture, and the other is our proposed Segment-Level Attention Mechanism (SAM).

Unlike most systems that output one probability score for an audio as a whole, we divide a 10-s audio clip into several segments. The output of our GCNN architecture is $\textbf{X}\in\mathbb{R}^{N\times C}$ and represents the probability for each class of each segment, where $N\in\mathbb{N}$ is the number of segments in an audio clip and $C\in\mathbb{N}$ is the number of predefined classes. The output of SAM is a vector $\textbf{W}\in\mathbb{R}^{N}$and represents the attention weight factor for each segment. Then we multiply $\textbf{X}$ with $\textbf{W}$ for each segment to obtain weighted segment scores. Those scores will be averaged among segments to get a vector $\textbf{Y}\in\mathbb{R}^{C}$ and then go through a softmax to represent the normalized probability for each class. The class with the largest probability is considered to be the classification result.
\begin{figure}[t]
\centerline{\includegraphics[width=0.8\columnwidth]{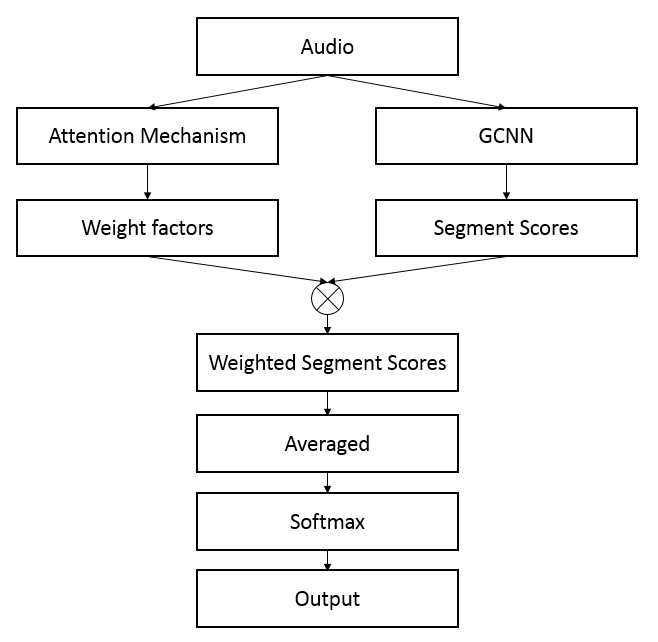}}
\caption{Overall architecture of proposed system}
\label{fig1}
\end{figure}
The detailed explanations of our proposed system will be included in the following parts of this section.

\subsection{Acoustic Feature}
We use fbank as the input of our system. Fbank is a two-dimensional time-frequency acoustic feature. It imitates the characteristics of human's ears and concentrates more on the low frequency components of audio signals. Compared with traditional MFCC feature, more original information can be kept in fbank and it has been widely used in deep learning. To extract fbank feature, each input audio is divided into 40ms frames with 50\% overlapping, and then 40 mel-scale filters are applied on the magnitude spectrum of each frame. Finally, we take logarithm on the amplitude and get fbank feature. As is mentioned in Section 1, the audio clips contain four channels, so our fbank feature contains four channels as well. In our work, four channels are fed into the system separately while training. And the averaged output score of four channels is used for evaluation.
\begin{figure*}[t]
\centerline{\includegraphics[width=1.8\columnwidth]{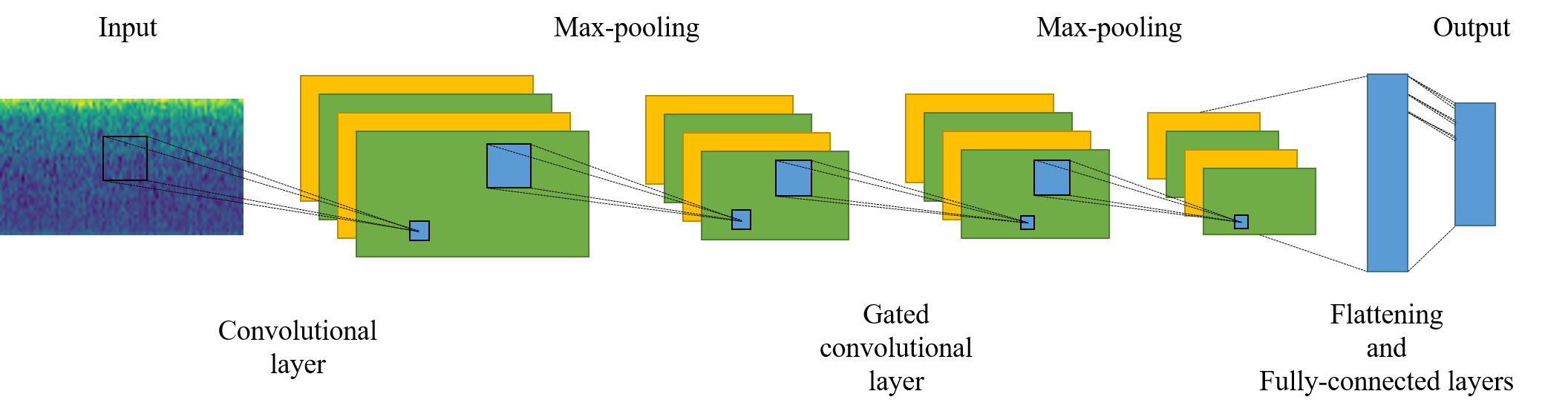}}
\caption{Overall architecture of gated convolutional neural network}
\label{fig}
\end{figure*}

\subsection{Gated Convolutional Neural Network}
Gated convolutional neural network was proposed by Dauphin et al. in \cite{gcnn} and has shown great power in machine translation, natural language processing. Our GCNN architecture consists of three main parts: 1) convolutional neural network (CNN), 2) gated convolutional neural network (GCNN), 3) feedforward neural network (FNN). And our overall architecture is shown in Figure 2.
\begin{figure}[ht]
  \centering
  \centerline{\includegraphics[width=0.8\columnwidth]{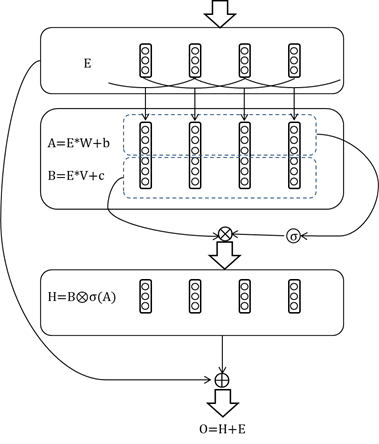}}
  \caption{ Gated convolutional neural network.}
  \label{fig:gcnn}
\end{figure}

Before being fed into GCNN architecture, the extracted fbank feature is normalized to zero mean and unit standard deviation (we call it global normalization, to distinct with the following time normalization).

Convolutional layers extract frequency features and connect features of adjacent frames. And the output of convolutional layer is followed by batch normalization \cite{BN}, a ReLU activation unit and a dropout layer \cite{Dropout}. Then a max-pooling layer is applied to keep the most important features.

The structure of gated convolutional neural network is illustrated in Figure 3.

In gated convolutional neural network, the output of convolutional layer is divided into two parts with the same size. The input of this structure is $E$ = [e$_1$, e$_2$, …, e$_n$], $E$ passes through a convolutional layer and the output is divided into $A$ and $B$. Then $A$ passes through sigmoid activation function and multiplies with $B$ by element-wise.
In order to enable stronger work, we add residual connections from the input $E$ to the output of this structure $H$. Residual network is introduced to avoid vanishing gradient problem\cite{ResNet}.

The specific formula is as follows:
\begin{equation}
  \label{eqn:gcnn_equation1}
    A=E*W+b,
\end{equation}
\begin{equation}
  \label{eqn:gcnn_equation2}
    B=E*V+c,
\end{equation}
\begin{equation}
  \label{eqn:gcnn_equation3}
    H=B\otimes\sigma(A),
\end{equation}
\begin{equation}
  \label{eqn:gcnn_equation4}
    O=H+E,
\end{equation}
where $W$, $V$ represent convolutional kernel values, and $b$, $c$ mean biases. $\otimes$ represents element-wise production. $\sigma(\cdot)$ is a sigmoid activation function.

The gated convolutional layer is also followed by batch normalization, a ReLU activation unit, a dropout layer and a max-pooling layer.

After the gated convolutional neural network, the features on multiple channels are flattened into frequency axis.

Then two fully-connected layers are used to combine extracted features and output nine scores for each segment. Our work differs from others in that we output scores for each segment while most researchers output scores for an audio as a whole. We intend to focus on those segments with more energy and ignore segments with less energy, which we call ``silence'' segments. That is why we propose a segment-level attention mechanism.

\subsection{Segment-Level Attention Mechanism}

\begin{figure}[t]
  \centering
  \centerline{\includegraphics[width=0.8\columnwidth]{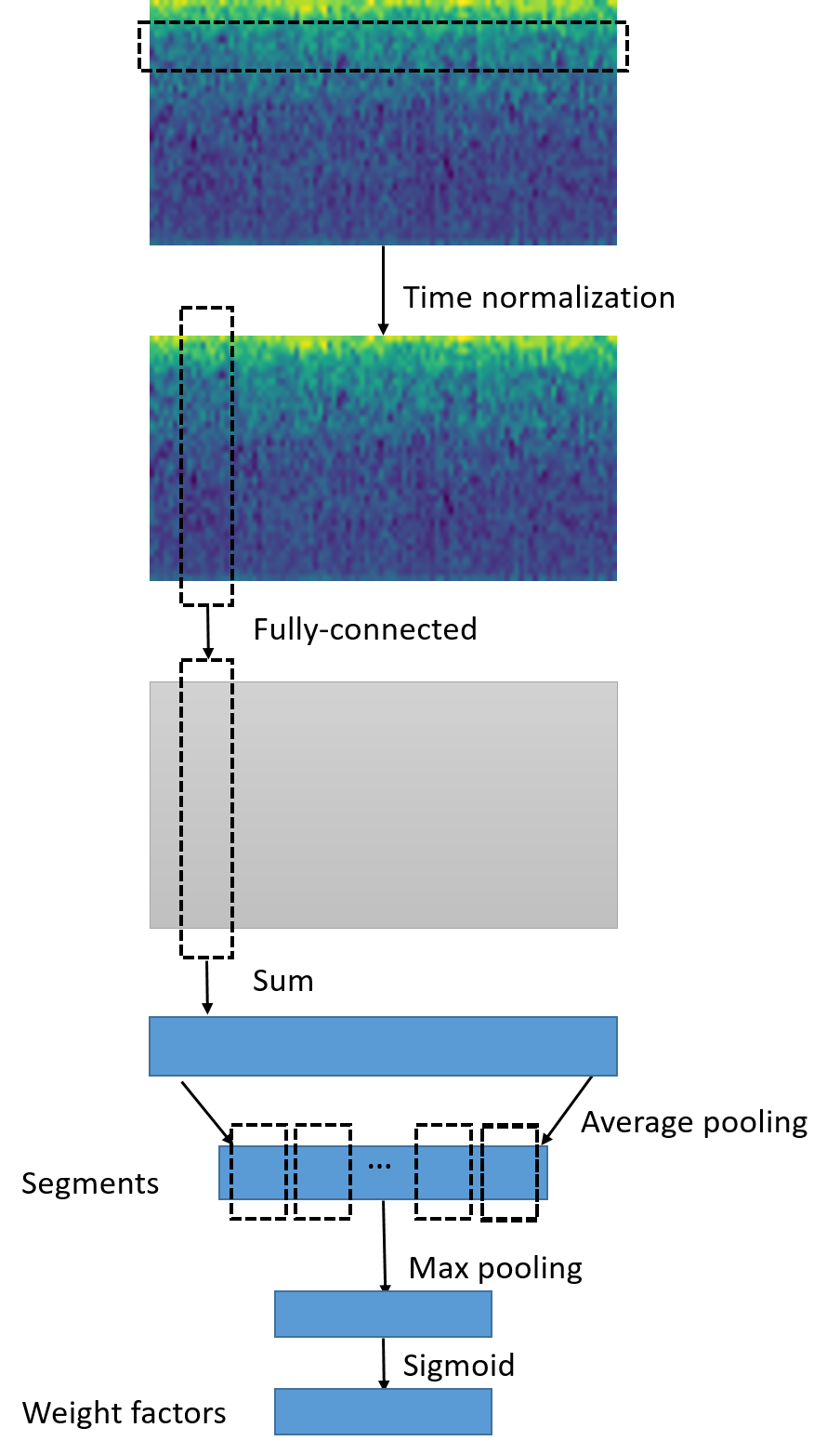}}
  \caption{Segment-Level Attention Mechanism.}
  \label{fig:sam}
\end{figure}

As mentioned in Section 1, attention mechanism was introduced to ignore irrelevant sounds such as background noise and silences in audio event classification. In DCASE 2018 task 5, an audio clip labeled as “cooking” may contain some segments of silences and we should not pay too much attention to those segments because audio clips labeled as other classes may also contain silences. Motivated by Xu et al. \cite{Xu}, we propose a segment-level attention mechanism. Our work differs from previous work in that we give our attention weight factors based on the characteristics of segments instead of frames.

The structure of segment-level attention mechanism is shown in Figure 4. The input of this structure is aforementioned fbank feature. Then it will be normalized along the time axis, which we call time normalization. The purpose of time normalization is to further differentiate the features among frames. 

A fully-connected layer is added to extract deeper features of frames. Like in the gated convolutional neural network, the fully-connected layer is followed by batch normalization, ReLU and dropout. Next, we calculate the sum along frequency axis. An average pooling layer is added to filter adjacent frames. Then a max-pooling layer is used to maintain the most important information of a segment. Finally, we use a sigmoid activation to limit the weight factors between 0 and 1. Based on our experiments, the duration of a segment is set to 1 second. Specific structure and hyperparameters will be illustrated in Section 3.

\subsection{Model Ensemble}
Model ensemble is a common strategy in machine learning. In our work, we propose a strategy of model ensemble.

During our experiments, we notice that “absence”, “other” and “working” are three sorts of activities that are often misclassfied with the others. So we train a model in particular to classify those three classes of activities. When our main system classifies an audio clip as any of the three classes, we will use the specially trained model for one more classification.

If an audio is classified as a class other than class 0, 4, 8 (“absence”, “other” and “working) by our first system, the output will be the final output. Otherwise, the audio will be fed into our second system. We denote the output of our first system as $\textbf{X}^{\rm \uppercase\expandafter{\romannumeral1}}\in \mathbb{R}^9$ and second system as $\textbf{X}^{\rm \uppercase\expandafter{\romannumeral2}}\in \mathbb{R}^3$. $\textbf{X}^{N}_i$ represents the output probability of \emph{i}-th class by the \emph{N}-th system, where $i\in [0,8]$ and \emph{N} is 1 or 2. Then the final output $\textbf{Y}\in \mathbb{R}^9$ of our ensemble system will be calculated according to the following algorithm. We calculate the sum of $\textbf{X}^{\rm \uppercase\expandafter{\romannumeral1}}_0,\textbf{X}^{\rm \uppercase\expandafter{\romannumeral1}}_4,\textbf{X}^ {\rm \uppercase\expandafter{\romannumeral1}}_8$ and redistribute them based on our second system output $\textbf{X}^ {\rm \uppercase\expandafter{\romannumeral2}}$.
\linespread{1.2}
\begin{algorithm}[!h]
	\caption{Model Ensemble$(\textbf{X}^{\rm \uppercase\expandafter{\romannumeral1}},\textbf{X}^{\rm \uppercase\expandafter{\romannumeral2}})$}
	\begin{algorithmic}%
		\STATE $O \gets argmax  \textbf{X}^{\rm \uppercase\expandafter{\romannumeral1}}$
        \STATE $\textbf{Y} \gets \textbf{X}^{\rm \uppercase\expandafter{\romannumeral1}}$
		\IF{$O==0\ or\ O==4\ or\ O==8$}
		\STATE $S \gets sum(\textbf{X}^{\rm \uppercase\expandafter{\romannumeral1}}_0,\textbf{X}^{\rm \uppercase\expandafter{\romannumeral1}}_4,\textbf{X}^{\rm \uppercase\expandafter{\romannumeral1}}_8)$
		\STATE $\textbf{Y}_{0,4,8} \gets S\textbf{X}^{\rm \uppercase\expandafter{\romannumeral2}}$
		\ENDIF
	\end{algorithmic}
\end{algorithm}

\section{EXPERIMENT, EVALUATION AND RESULTS}
\subsection{Experiment setup}
Our model is trained using Adam \cite{Adam} for gradient based optimization. Cross-entropy is used as the loss function. And the structure of our system is shown in Table 2 and Table 3 along with parameters. The initial learning rate is 0.001 and  the batch size is 256$\times$4 channels because each channel is considered as a different sample for training. We train the classifiers for 300 epochs.

We select 5\% of the testing data as validation dataset and choose models which result in the best accuracy on the validation dataset for final evaluation. In the evaluation process, the outputs of 4-channel acoustics are averaged to get the final posterior probability.

\begin{table}[h]
  \centering
  \caption{Model structure and parameters of gated convolutional neural network}
    \setlength{\abovetopsep}{0.5ex}
    \setlength{\belowrulesep}{0pt}
    \setlength{\aboverulesep}{0pt}
    \renewcommand\arraystretch{1.2}
    \begin{tabular}{c|c}
    \toprule
    Input 40$\times$501$\times$1&	Output size\\
    \midrule
    Conv (padding: valid, kernel: [40, 5, 64])& 1, 497, 64 \\
    BN-ReLU-Dropout(0.2)& 1, 497, 64\\
    1$\times$5 Max-Pooling(padding: valid)& 1, 99, 64\\
    Gated Conv (padding: same, kernel: [1, 3, 128])& 1, 99, 64\\
    BN-ReLU-Dropout(0.2) & 1, 99, 64\\
    1$\times$10 Max-Pooling(padding: same) & 1, 10 64\\
    Feature Flattening& 10, 64\\
    Fully-connected(unit num: 64) -ReLU-Dropout(0.2)& 10, 64\\
    Fully-connected(unit num: 9)& 10, 9\\
    \bottomrule
    \end{tabular}
  \label{table2}
\end{table}

\begin{table}[h]
  \centering
  \caption{Model structure and parameters of segment-level attention mechanism}
    \setlength{\abovetopsep}{0.5ex}
    \setlength{\belowrulesep}{0pt}
    \setlength{\aboverulesep}{0pt}
    \renewcommand\arraystretch{1.2}
    \begin{tabular}{c|c}
    \toprule
    Input 40$\times$501$\times$1&	Output size\\
    \midrule
   	Fully-connected(unit num: 40)& 40, 501, 1 \\
    BN-ReLU-Dropout(0.2)& 40, 501, 1\\
    Sum along frequency axis& 1, 501, 1\\
    1$\times$5 Average-Pooling(padding: same)& 1, 100, 1\\
    1$\times$10 Max-Pooling(padding: same) & 1, 10, 1\\
    Squeeze& 10\\
    Sigmoid& 10\\
    \bottomrule
    \end{tabular}
  \label{table3}
\end{table}

\subsection{Evaluation Metric}
The official evaluation metric for DCASE 2018 challenge task 5 is macro-averaged F1-score. F1-score is a measure of a test's accuracy and it is the harmonic average of precision and recall. Macro-averaged means that F1-score is calculated for each class separately and averaged over all classes. For this task, a full 10s multi-channel audio is considered to be one sample.

\subsection{Results}
We examine the following configurations:

(1) CNN: Convolutional neural network as baseline system;

(2) SAM-CNN: Convolutional neural network with our proposed segment-level attention mechanism;

(3) GCNN: Gated convolutional neural network;

(4) SAM-GCNN: Gated convolutional neural network with our proposed segment-level attention mechanism;

(5) Ensemble: Gated convolutional neural network with our proposed segment-level attention mechanism and model ensemble.

\begin{table}[h]
  \centering
  \caption{Macro-averaged F1-score of multiple systems on 4 folds}
  \begin{tabular}{cccccc}
    \toprule
    System&	Fold1&	Fold2&	Fold3&	Fold4&	Average\\
    \midrule
    CNN&	81.92\%& 	82.58\%& 	83.26\%& 	87.29\%& 	83.76\%\\
    GCNN&	85.58\%& 	84.22\%& 	86.36\%& 	88.83\%& 	86.25\%\\
    SAM-CNN&	83.68\%& 	82.26\%& 	84.56\%& 	88.09\%& 	84.65\%\\
    SAM-GCNN&	88.49\%& 	86.81\%& 	86.51\%& 	90.52\%& 	88.08\%\\
    Ensemble&	89.62\%& 	88.11\%& 	87.95\%& 	91.63\%& 	89.33\%\\
    \bottomrule
    \end{tabular}
  \label{table4}
\end{table}
As shown in Table 4, the macro-averaged F-1 score of GCNN is 2.49\% higher than CNN. And our proposed segment-level attention mechanism can improve the classification performance of both CNN and GCNN.

Moreover, our proposed ensemble strategy can outperform previous systems and achieve 89.33\% F1-score. Confusion matrix before and after ensemble is shown in Figure 5. On the left is the confusion matrix of SAM-GCNN, and on the right is the confusion matrix of SAM-GCNN with model ensemble. The element in the \emph{i}-th row and \emph{j}-th column of this matrix represents the amount of audio clips that belong to class \emph{i} and are classified as class \emph{j}, so the elements on the diagonal represent the number of correctly classified audio clips. We can find that the number of correctly classified audio clips has increased after ensemble, especially for ``absence", ``other" and ``working", showing that our model ensemble method does work.

The class-wise performance of our final model is shown in Table 5.

\begin{table}[h]
  \renewcommand\tabcolsep{5.0pt}
  \centering
  \caption{Class-wise performance of proposed model}
  \begin{tabular}{cccccc}
    \toprule
        &  fold1&	fold2&	fold3&	fold4&	Average\\
    \midrule
    Absence& 94.43\%& 92.99\%& 	93.15\%& 	94.93\%& 	93.88\%\\
    Cooking& 95.92\%& 94.26\%& 	93.75\%& 	96.49\%& 	95.10\%\\
    Dishwashing& 87.45\%& 81.22\%& 	81.81\%& 	83.87\%& 	83.59\%\\
    Eating&	89.35\%& 89.66\%& 87.73\%& 	90.56\%& 	89.33\%\\
    Other&	52.28\%&  53.51\%& 54.61\%&  67.15\%& 	56.89\%\\
    Social activity& 97.83\%& 95.85\%& 	94.38\%& 	98.50\%& 	96.64\%\\
    Vacuum cleaning& 99.99\%& 99.81\%& 	100.00\%& 	100.00\%& 	99.95\%\\
    Watching TV& 99.55\%& 99.86\%& 99.42\%& 	99.91\%& 	99.69\%\\
    Working& 89.82\%& 85.85\%& 86.68\%& 	93.22\%& 	88.89\%\\
    Macro-Average& 89.62\%& 88.11\%& 	87.95\%& 	91.63\%& 	89.33\%\\
    \bottomrule
    \end{tabular}
  \label{table5}
\end{table}

To better evaluate our work, we compare the performance of proposed model with the top-2 ranked teams in DCASE 2018 Challenge Task 5 and the official baseline system in Table 6. Both of the top-2 teams adopted complex methods of pre-processing, data augmentation and model ensemble. We can achieve equivalent performance without any data augmentation. And our system outperforms the official baseline significantly.

\begin{table}[h]
  \centering
  \caption{Comparison with state-of-the-art works}
    \begin{tabular}{cc}
    \toprule
	&	Averaged F1-score\\
    \midrule
    Proposed&	89.3\%\\
    InouetMilk\cite{IBM}& 90.0\%\\
    HITfweight\cite{HIT}&	89.8\%\\
    Official Baseline& 84.5\%\\
    \bottomrule
    \end{tabular}
  \label{table6}
\end{table}

\begin{figure*}[t]
\centerline{\includegraphics[width=1.8\columnwidth]{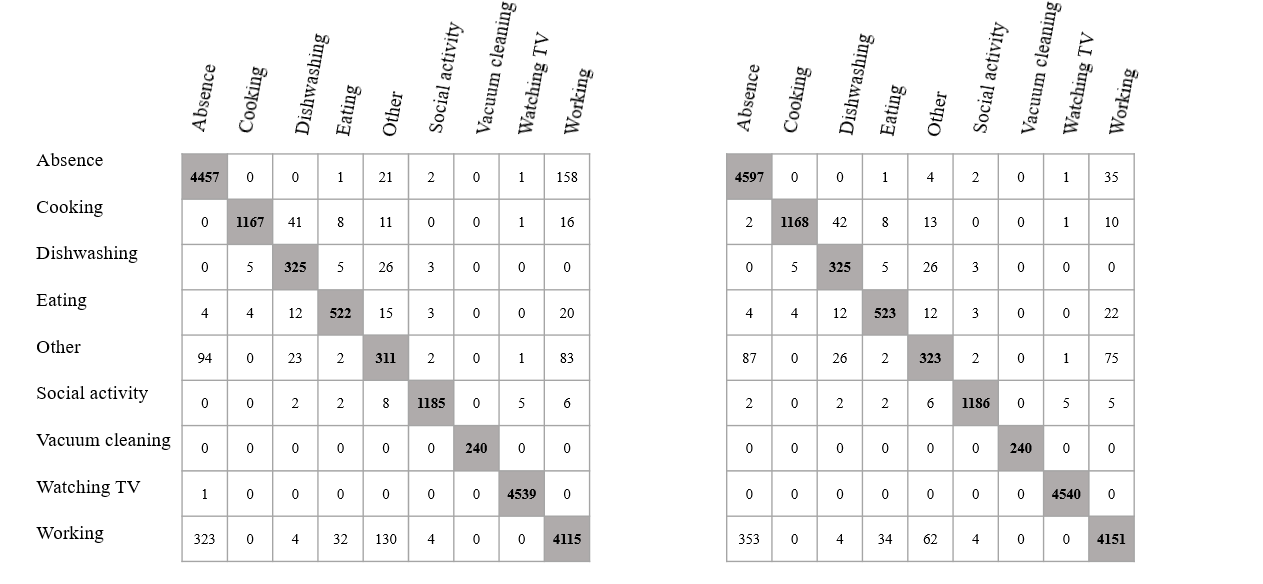}}
\caption{Confusion matrix before and after ensemble on fold4.}

\label{fig5}
\end{figure*}

\section{Conclusion}
In this paper, we have introduced our work and the results show that the performance of our proposed system is significantly superior to that of the baseline. Our proposed segment-level attention mechanism improves the performance of both CNN and GCNN architecture. Furthermore, by using model ensemble, we have achieved competitive performance on the development dataset of DCASE 2018 task 5. Note that both the top two teams of this task utilized complex methods of data augmentation and model ensemble. Our system can achieve equivalent performance without data augmentation, which shows that our proposed attention mechanism can contribute a lot to home activity monitoring. Since the ground truth labels of evaluation dataset of DCASE 2018 challenge have not been published yet, future work needs to be done for further evaluation.

\bibliographystyle{unsrt}

%
\end{document}